\documentclass[amsfonts,amssymb,prl, superscriptaddress, notitlepage, twocolumn, nofootinbib]{revtex4-2}

\usepackage{amsmath}    
\usepackage{graphicx}   
\usepackage{verbatim}   
\usepackage{color}      
\usepackage{hyperref}   
\usepackage[english]{babel}
\usepackage{ucs}
\usepackage[utf8x]{inputenc}
\usepackage{subfigure}
\usepackage[autostyle]{csquotes}
\usepackage{float}
\usepackage{textcomp}

\newcommand{\Cov}{\operatorname{Cov}}
\DeclareMathOperator{\E}{\mathbb{E}} 

\newcommand{\bc}[2]{$\beta$#1cell#2}
\newcommand{\lra}[1]{\langle{}#1\rangle}

\begin{document}

\title{Autopoietic influence hierarchies in pancreatic \bc{ }{s}}

\author{Dean Koro\v{s}ak}
\affiliation{Institute of Physiology, Faculty of Medicine, University of Maribor, 2000 Maribor, Slovenia}
\affiliation{Faculty of Civil Engineering, Transportation Engineering and Architecture, University of Maribor, 2000 Maribor, Slovenia}

\author{Marko Jusup}
\affiliation{Tokyo Tech World Research Hub Initiative (WRHI), Institute of Innovative Research, Tokyo Institute of Technology, Tokyo 152-8552, Japan}

\author{Boris Podobnik}
\affiliation{Faculty of Civil Engineering, University of Rijeka, 51000 Rijeka, Croatia}
\affiliation{Center for Polymer Studies, Boston University, Boston MA 02215, USA}
\affiliation{Zagreb School of Economics and Management, 10000 Zagreb, Croatia}
\affiliation{Luxembourg School of Business, 2453 Luxembourg, Luxembourg}
\affiliation{Faculty of Information Studies in Novo mesto, 8000 Novo Mesto, Slovenia}

\author{Andra\v{z} Sto\v{z}er}
\affiliation{Institute of Physiology, Faculty of Medicine, University of Maribor, 2000 Maribor, Slovenia}

\author{Jurij Dolen\v{s}ek}
\affiliation{Institute of Physiology, Faculty of Medicine, University of Maribor, 2000 Maribor, Slovenia}
\affiliation{Faculty of Natural Sciences and Mathematics, University of Maribor, 2000 Maribor, Slovenia}

\author{Petter Holme}
\thanks{Corresponding author: holme.p.aa@m.titech.ac.jp}
\affiliation{Tokyo Tech World Research Hub Initiative (WRHI), Institute of Innovative Research, Tokyo Institute of Technology, Tokyo 152-8552, Japan}

\author{Marjan Slak Rupnik}
\thanks{\mbox{Corresponding author: marjan.slakrupnik@meduniwien.ac.at}}
\affiliation{Institute of Physiology, Faculty of Medicine, University of Maribor, 2000 Maribor, Slovenia}
\affiliation{Center for Physiology and Pharmacology, Medical University of Vienna, 1090 Vienna, Austria}
\affiliation{Alma Mater Europaea -- European Center Maribor, 2000 Maribor, Slovenia}

\date{\today}

\begin{abstract}
\bc{ }{s} are biologically essential for humans and other vertebrates. Because their functionality arises from cell-cell interactions, they are also a model system for collective organization among cells. There are currently two contradictory pictures of this organization: the hub-cell idea pointing at leaders who coordinate the others, and the electrophysiological theory describing all cells as equal. We use new data and computational modeling to reconcile these pictures. We find via a network representation of interacting \bc{ }{s} that leaders emerge naturally (confirming the hub-cell idea), yet all cells can take the hub role following a perturbation (in line with electrophysiology).
\iffalse
A useful network representation of complex systems is via pairwise cross-correlations between a system's constituents. Recently, attempts have been made at representing pancreatic \bc{-}{} collectives as networks, giving birth to the concept of hub cells, which contradicts established electrophysiological theory. Here, we reconcile the hub-cell idea with electrophysiology using empirical and computational methods. While the former reveals the functional importance of hubs, the latter show that hubs are an emergent property of communication among identical peers. Thus, individual cells are perfectly capable of playing the hub role, but in line with electrophysiology, they do not have to be genetically predisposed to do so. This points to an ultra-robust architecture of \bc{-}{} collectives and underpins some of their vital features.
\fi
\end{abstract}

\maketitle 
 
%%%%% beta cells collective and coupling %%%%%%%%%%
%\section{Introduction}

The importance of the nutrient-sensing and insulin-secreting \bc{ }{s} in vertebrates is hard to overstate. These cells reside in pancreatic islets, where they extensively communicate with each other and their environment~\cite{pipeleers1987biosociology}. The intercellular communication, in particular, serves to coordinate and synchronize cellular operations through which insulin is released in proportion to stimulation and metabolic requirements~\cite{pipeleers1982glucose, korosak2018collective}. The delicate nature of this task is reflected in the continual need to prevent oversecretion, and subsequent hypoglycemia~\cite{kolb2020insulin}, despite intracellular stores holding sufficient insulin to exceed the lethal dose by orders of magnitude if released at once.

On a microscopic scale, \bc{ }{s} are electrically and metabolically coupled via gap junctions built from the protein connexin-36 (Cx36)~\cite{meda2018gap}. The key role of such coupling is seen in the fact that the loss of Cx36 channels has a detrimental impact on \bc{-}{} cooperation~\cite{ravier2005loss, speier2007cx36}, leading to uncoordinated plasma depolarizations, the desynchronization of calcium signals, and increases in basal insulin release. Sufficient Cx36 coupling, by contrast, curbs the \bc{-}{} intrinsic heterogeneity in glucose sensitivity~\cite{benninger2018new}, and in mouse islets, limits the threshold for stimulatory glucose concentration~\cite{speier2007cx36} to a narrow band around $\approx$7\,mM. Metaphorically, coupled \bc{ }{s} are like individual soldiers who fall in line when the communication channels between them are open. A typical \bc{ }{} is coupled to between six and eight neighbors~\cite{zhang2008cell, skelin2017triggering}. According to electrophysiological theory~\cite{satin2020take}, single cells lack mechanisms to become pacemakers beyond their immediate neighborhood. In terms of our metaphor, there are no generals in the army.

The above-mentioned lateral organization is diametrically opposed to the picture by functional studies of a large number of communicating \bc{ }{s} (referred to as \bc{-}{} collectives). For example, intrinsic cellular heterogeneity in glucose sensing~\cite{benninger2018new}, heterogeneous gap-junction coupling~\cite{farnsworth2014fluorescence}, and extensive paracrine signaling~\cite{caicedo2013paracrine} all contribute to an islet-wide complicated cytoplasmic Ca$^{2+}$ dynamics. This dynamics of the Ca$^{2+}$ concentration in the cytosol~\cite{berridge2000versatility, colecraft2020research} is a key insulin-secretion driver~\cite{idevall2020metabolic}. It has recently become possible to record the Ca$^{2+}$ dynamics with a great spatial and temporal precision using the functional multicellular confocal imaging~\cite{stozer2013glucose, dolensek2020glucose}. The high data content of such imaging enables mapping the functional organization of a \bc{-}{} collective onto a complex network such that a pair of cells (i.e., network nodes) is linked if the cross-correlation between the corresponding Ca$^{2+}$ signals is large enough~\cite{hodson2012existence, stozer2013functional, cherubini2015role, johnston2016beta, gosak2018network, salem2019leader}.

Complex-network representations of the functional organization of \bc{-}{} collectives (i.e., functional networks) that arise from calcium cross-correlations in intact pancreatic islets reveal a modular structure~\cite{markovic2015progressive} intertwined with small-world~\cite{rutter2013minireview, stozer2013functional} properties. Of particular interest is that some studies~\cite{johnston2016beta, salem2019leader} point to small subsets of highly active \bc{ }{s} whose connectedness and impact on synchronization across islets make these cells candidate leaders. Because of their large degree, we call such candidate leaders \textit{hub cells}. The proponents of the hub-cell idea list many reasons~\cite{rutter2020comment} why electrophysiological measurements may have missed detecting hubs. Overall, functional networks suggest that our army of \bc{ }{s} is led by a few generals.

We will hereafter show that the hub-cell idea, rather than contradicting, actually complements electrophysiology. Based on a remarkable agreement between empirical and computational findings, we will argue in favor of an autopoietic influence hierarchy among \bc{ }{s}. Leader cells emerge through cell-cell communication, and thus need not be genetically predisposed for leadership. In the language of the army metaphor, generals do lead, but not by birthright; they get promoted by their peers. Leader cells, furthermore, remain under the radar of electrophysiological measurements by communicating with immediate neighbors just as any other cell would. A direct implication is that a more-or-less arbitrary cell could emerge as a leader, which in turn makes for an ultra robust architecture of \bc{-}{} collectives. Additional wide-reaching implications underpin some of the vital \bc{-}{} features.

%%%%%%%% methods %%%%%%%%%%%%%%
\section{Methods}

We based our empirical analyses on a dataset obtained via Ca$^{2+}$ imaging of an acutely prepared pancreatic tissue slice~\cite{speier2003novel, stozer2013glucose} comprising a rodent oval-shaped islet (approx.\ dimensions: 370\,{\textmu}m$\times$200\,{\textmu}m). Ca$^{2+}$ signals were recorded using a functional multi-cellular imaging technique at 10\,Hz and 256$\times$256 pixel resolution in 8-bit grayscale color depth. The freely downloadable dataset~\cite{Jusup2020} consisted of 65,536 Ca$^{2+}$ signals, each with 23,873 data points. Our focus was on fast oscillations, which is why all signals were detrended and standardized before use~\cite{podobnik2020beta}.

For the purpose of constructing empirical functional networks, we randomly picked $N=100$ signals, denoted $x_i(t)$, from the dataset and computed cross-correlations $c_{ij} = [x_i x_j]$, where $[\cdot]$ is a time-averaging operator. The number of data points in each signal was $L=3 \times 10^3$, corresponding to 5-minute exposures of the pancreatic tissue slice to the stimulating glucose concentration of 8\,mM. We link two signals $i$ and $j$ in the network if $c_{ij} > c_{0}$, where $c_{0}$ is a threshold selected to give a mean degree $\lra{k} = 10$.
%%% describe previous RMT results, frontiers paper, arguments for statistically significant correlations %%%%%
We have previously shown~\cite{slak2019random} that the bulk cross-correlation spectrum of typical Ca$^{2+}$ signals follows the predictions of random matrix theory, but the states outside the bulk-spectrum edges carry biological information. The delocalized states corresponding to the largest eigenvalues of the cross-correlation matrix, in particular, were found to harbor contributions from all Ca$^{2+}$ signals, thus revealing an islet-wide collective mode~\cite{plerou2002random}.

\begin{figure}
\centering
\includegraphics[width=\linewidth]{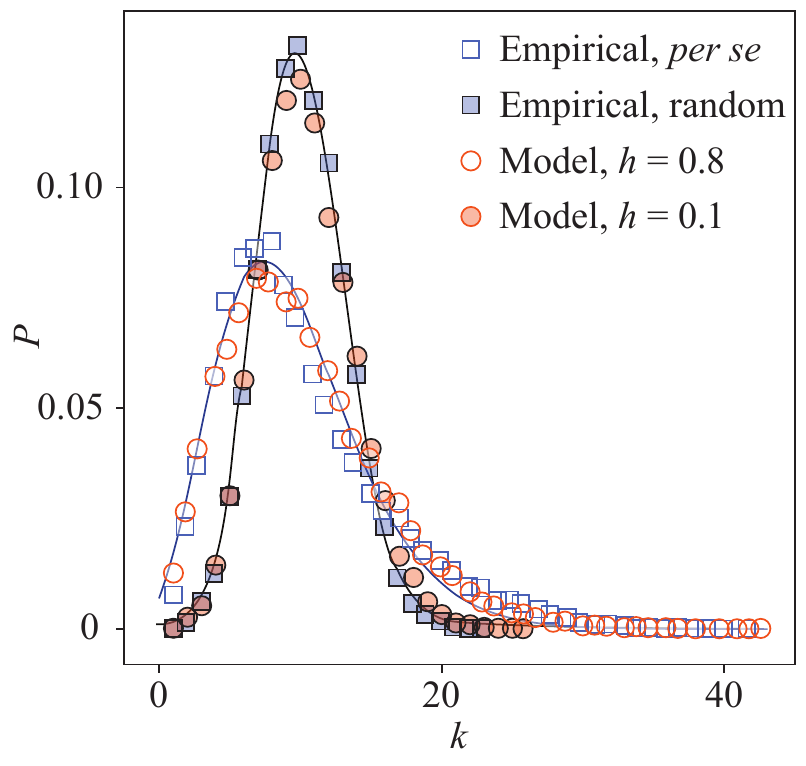}
\caption{\label{fig:fig1}  Functional networks of \bc{ }{s} contain high-degree hubs. The plot shows the configuration-ensemble degree distribution, $P$, as a function of the node degree, $k$.
}
\end{figure}

The computational aspects of the study comprised constructing functional networks from simulated Ca$^{2+}$ signals. Simulations were based on our model~\cite{podobnik2020beta} that had been found to mimic empirical signals closely. Compared to typical electrophysiological modeling by means of differential-algebraic systems~\cite{cherubini2015role}, the model's structure is very simple. Each of the $N=100$ nodes represents a \bc{ }{} arranged in a random regular network with the degree distribution $P(k) = \delta (k - k_0)$ such that trivially the average node degree is $\lra{k}=k_0$. Nodes can change their binary state, indicating the calcium activity of individual \bc{ }{s}, from active to inactive or vice versa in two ways. Internal activation is controlled by a forcing parameter, $f$, that is interpretable as the inverse of the glucose concentration to which \bc{ }{s} are exposed. External activation is controlled by the state of the $k_0$ nearest neighbors in conjunction with the coupling strength, $r$, that is interpretable as the intensity of gap-junctional ion exchange. Functional networks were constructed from simulated Ca$^{2+}$ signals in exactly the same way as from empirical ones.

%%%%%% results %%%%%%%%%
\section{Results}

The degree distributions of empirical and simulated functional networks are practically the same (Fig.~\ref{fig:fig1}). We distinguish between two cases. Open symbols pertain to functional networks extracted from empirical data \textit{per se} (squares), and from simulations with strong coupling, $h=0.8$ (circles). In this case, the obtained degree distributions closely follow a negative binomial distribution $P(k) = \binom{k+r-1}{k}p^k (1-p)^r$, where $p =\lra{k}/(r + \lra{k})\approx{}0.77$ and $r=3$ is a numerical parameter. The opposite case, shown using filled symbols, consists of functional networks extracted upon randomizing empirical data (squares) and running simulations with weak coupling, $h=0.1$ (circles). Both data randomization and weak coupling should produce random functional networks, what indeed transpires given that the obtained degree distributions closely follow a binomial distribution $P(k) = \binom{N}{k}p^k (1-p)^{N-k}$ with $p = \lra{k}/N$. The binomial distribution in turn converges to a Poisson distribution $P(k) = \lra{k}^k\exp(-\lra{k})/k!$ in the large network limit $N\to\infty$ when $\lra{k}$ is fixed.

% In this limit the negative binomial distribution also tends to Poisson distribution for $r\to\infty$, and for large $k$ to gamma distribution $P(k) = q^{-r} k^{r-1}\exp(-k/q)/\Gamma(r)$ where $qr =\langle k\rangle$.

\begin{figure}
\centering
\includegraphics[width=\linewidth]{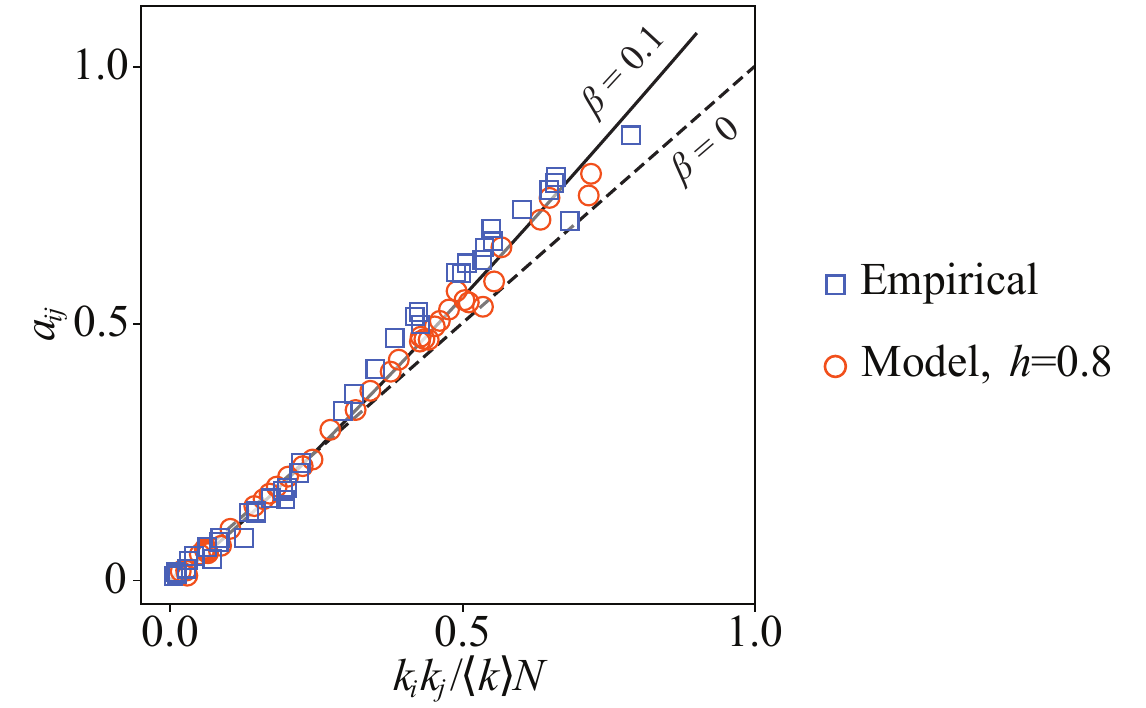}
\caption{\label{fig:fig2} \bc{-}{} functional networks exhibit positive assortative mixing. The plot compares the elements $a_{ij}$ of the configuration-ensemble adjacency matrix against the normalized degree product, $(k_ik_j/\lra{k}N)^{1+\beta}$, for two different values of the $\beta$ exponent, $\beta=0$ and $\beta=0.1$.}
\end{figure}

Aside from degree distributions, another informative way of characterizing functional networks is degree correlations, also known as degree assortativity. Ref.~\cite{johnson2010entropic} explains that the elements $a_{ij}$ of a configuration-ensemble adjacency matrix should satisfy the strictly linear relationship $a_{ij} = k_i k_j/\lra{k} N$ in zero-assortativity situations. Otherwise, $a_{ij}=(k_i k_j/\lra{k} N)^{1+\beta}$, where $\beta<0$ ($\beta>0$) indicates negative (positive) assortative mixing. Plots of $a_{ij}$ against $k_i k_j/\lra{k} N$ (Fig.~\ref{fig:fig2}), while once again revealing an agreement between empirical and simulated functional networks, primarily show that $\beta\approx0.1$. The degree assortativity of the functional networks is therefore positive, meaning that network nodes are preferentially linked to other nodes of similar degree.

\begin{figure}
\centering
\includegraphics[width=\linewidth]{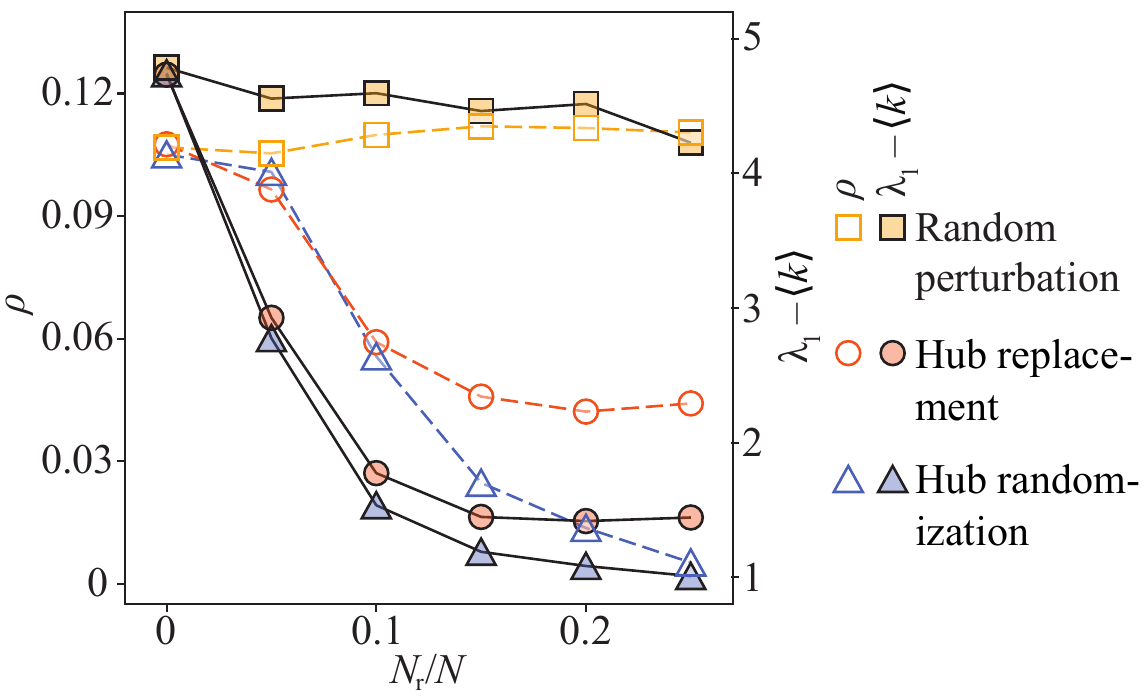}
\caption{\label{fig:fig3} Assortativity and the largest adjacency-matrix eigenvalue of functional networks are robust to random perturbations, but not to perturbations targeting hubs. The plot shows assortativity, $\rho$, and the largest eigenvalue of the configuration-ensemble adjacency matrix, $\lambda_1$, (from which the average node degree, $\lra{k}$, has been subtracted) both as the functions of the fraction of perturbed nodes, $N_\mathrm{r}/N$.} 
\end{figure}

\begin{figure*}
\centering
\includegraphics[width=\linewidth]{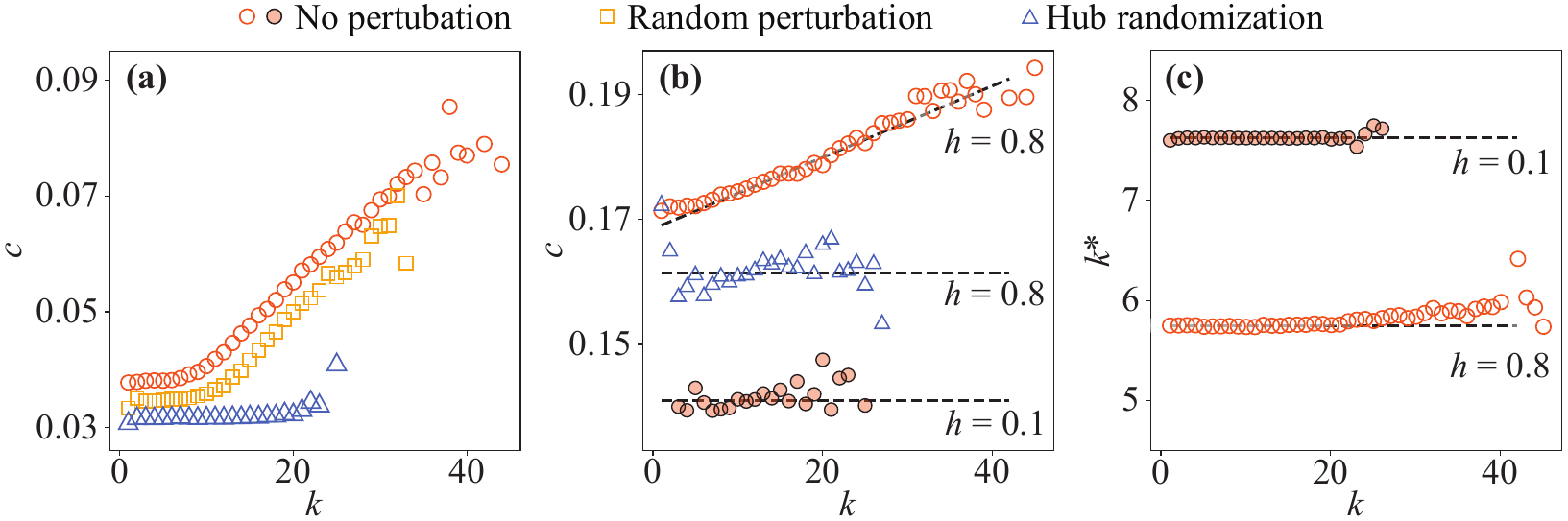}
\caption{\label{fig:fig4} Hub cells carry most of the cross-correlation content of a \bc{-}{} collective. (a) Empirical neighborhood-wide average cross-correlation of nodes, $c$, as a function of the node degree, $k$. Open circles, squares, and triangles respectively denote no perturbation, random perturbation, and hub randomization. (b) Same as (a), but simulated. Open circles and triangles respectively denote no perturbation and random perturbation when coupling, $h=0.8$ is strong, whereas filled circles denote no perturbation when coupling, $h=0.1$, is weak. (c) The simulated average number of active neighbors, $k^*$, as a function of the node degree. Open and filled circles respectively denote the strong, $h=0.8$, and the weak coupling, $h=0.1$.}
\end{figure*}

The results so far allow us to estimate the assortativity coefficient of functional networks in an alternative way, which we can then compare with the definition~\cite{newman2002assortative, qu2015effects}, that is, with the degree correlation coefficient between pairs of linked nodes $\rho = (\lra{k}\lra{k^2 k_\mathrm{nn}} - \lra{k^2}^2) /
(\lra{k}\lra{k^3} - \lra{k^2}^2)$, where $k_\mathrm{nn}^i = k_i^{-1}\sum_j a_{ij}k_j$ is the average degree of the $i$th node's nearest neighbors. Using the exponent $\beta\approx0.1\ll1$ and the fact that the aforementioned negative binomial distribution converges to a gamma distribution $P(k) = q^{-r} k^{r-1}\exp(-k/q)/\Gamma(r)$ for large $q=\lra{k}/r$, the assortativity coefficient is $\rho\approx\beta{}q^{2\beta}\approx{}0.11$ (see Ref.~\cite{johnson2010entropic}). Following the definition of $\rho$ indeed yields a very similar value (Fig.~\ref{fig:fig3}; open symbols for $N_\mathrm{r}/N=0$).

We furthermore tested how two different perturbations~\cite{albert2000error}, signal randomization and replacement, affect the assortativity coefficient. Here, randomization and replacement respectively mean that a fraction $N_\mathrm{r}/N$ of signals used in the construction of the original configuration ensemble were either randomly reshuffled or replaced with other random signals from the broader dataset. We found that assortativity of functional networks is robust to random perturbations, but not to perturbations targeting hubs (Fig.~\ref{fig:fig3}; open symbols for $N_\mathrm{r}/N>0$). It is therefore hubs that preferentially link to other hubs, which emphasizes their importance in the overall structure of functional networks. Specifically, we are seeing the evidence of a purposeful correlating mechanisms at work among \bc{-}{s} because otherwise heterogeneous networks are expected to be disassortative~\cite{johnson2010entropic, qu2015effects}.

Another measure characterizing a network's structure is the largest eigenvalue of the adjacency matrix, $\lambda_1$. Similarly as with the assortativity coefficient, the value of $\lambda_1$ is robust to random perturbations, but not to perturbations targeting hubs (Fig.~\ref{fig:fig3}). As progressively more hubs get perturbed, the maximum adjacency-matrix eigenvalue falls to $\lambda_1=1+\lra{k}$, which is the value characteristic of random networks with a Poisson degree distribution. This last result reaffirms the importance of hubs for the structure of functional networks and, in turn, hints at an interesting connection with recent experimental findings that generated much excitement.

Experiments show~\cite{johnston2016beta, salem2019leader} that perturbing hub cells can quickly decorrelate a \bc{-}{} collective. This implies that hubs carry most of the collective's cross-correlation content, which could be tested using our empirical and simulated Ca$^{2+}$ signals alike. To this end, we employed the pairwise cross-correlations $c_{ij}$ to define $c_i = k_i^{-1} {\sum_j a_{ij}c_{ij}}$, that is, the $i$th node's neighborhood-wide average cross-correlation. By pairing the $c_i$ values with the corresponding node degree $k_i$, we obtained a function $c(k)$. If the \bc{-}{} collective's cross-correlation content were equally distributed among all nodes, then we should see $c(k)=\mathrm{const.}$, whereas if hubs carried most of the content, then $c(k)$ should be an increasing function of $k$. We find that the latter is true both for empirical (Fig.~\ref{fig:fig4}a) and simulated (Fig.~\ref{fig:fig4}b) Ca$^{2+}$ signals. As with assortativity $\rho$ and the largest eigenvalue $\lambda_1$, random perturbations have little bearing on the function $c(k)$, but targeting high-degree nodes causes decorrelation. With the help of the dynamical network model, we see that perturbing about 10\,\% of hubs is very similar to running simulations with a weak coupling of $h=0.1$ (Fig.~\ref{fig:fig4}b).

The model additionally allows us to examine whether the node-degree dependence of the function $c(k)$ could be due to hubs somehow communicating with more of their neighbors than lower-degree nodes. The number of active neighbors $k^*$ is, as expected, independent of the degree $k$ when the coupling is weak, $h=0.1$ (Fig.~\ref{fig:fig4}c). The same, however, approximately holds even when the coupling is strong, $h=0.8$ (Fig.~\ref{fig:fig4}c). Because, in the model, all nodes are equal and none of them communicate with an unusual number of neighbors, we are left with a conclusion that the emergence of hubs in functional networks is endogenous to cell-cell communication.

%%%%%% discussion %%%%%%%%%
\section{Discussion}

Herein, we used a combination of empirical and computational methods to shed new light on, among others, a fundamental tension between electrophysiological theory and the hub-cell idea as it pertains to pancreatic \bc{ }{s}. We started by constructing functional \bc{-}{} networks from empirical Ca$^{2+}$ signals, and after that  estimated the configuration ensemble~\cite{molloy1995critical, nadakuditi2013spectra} of such networks.
%that is, the expected proportion of nodes of a given degree, as well as the probabilities that two connected nodes have given degrees.
Meanwhile, we also simulated fast \bc{-}{} activity using a dynamical network model~\cite{majdandzic2014spontaneous, podobnik2017biological} whose outputs faithfully mimic the properties of said empirical Ca$^{2+}$ signals~\cite{podobnik2020beta}. Upon repeating the construction of functional networks, but now from simulated signals, we found remarkable quantitative agreement between the results.

Irrespective of the signal origin, empirical or simulated, the properties of functional networks---such as assortativity or the largest eigenvalue of the configuration-ensemble adjacency matrix---all critically depend on the presence of highly connected (i.e., hub) nodes. Our results thus support the hub-cell idea. A similar heterogeneous organization of functional networks has also been discovered in the collectives of chemosensing cells~\cite{sun2013network, potter2016communication} communicating via gap junctions.

The support for the hub-cell idea would be a strong blow to electrophysiological theory if nodes in the dynamical network model had any distinctive properties, for example, different degrees or an internal structure. In the model, however, all nodes are completely identical, upholding the ideas of electrophysiological theory that no cell is predisposed for leadership. Leader cells, in fact, emerge through cell-cell communication as evidenced by tracing the degree distributions of functional networks to the coupling strength in the dynamical network model. In the language of the army metaphor, generals do lead, but not by birthright; they get promoted by their peers. Our findings thus not only reconcile the concept of hubs with electrophysiological theory, but also point to an extremely robust architecture of collectives.

Before expanding on the last remark about the robustness of \bc{-}{} collectives, let us examine a physiological advantage provided by the heterogeneity of functional networks. Cells have been shown to use cell-cell distance to optimize their sensing precision~\cite{fancher2017fundamental}. In compact microorgans such as pancreatic islets, cells cannot easily adjust mutual distances, but could instead use functional-network heterogeneity to sharpen their collective response. Let $\lambda_1$ be the largest adjacency-matrix eigenvalue and $\mathbf{v}^1$ the corresponding eigenvector~\cite{farkas2001spectra, plerou2002random}. Because the components of the latter vector quantify the importance of network nodes, we assume that $y_i = \sum_{j\in\Omega_i} v_j^1 x_j = \sum_{j} a_{ij} v_j^1 x_j$ represents the signal that node $i$ integrates from the nearest neighbors $j\in\Omega_i$, while all other signals are treated as constant-variance noise. Under this assumption, the signal-to-noise ratio of sensing cells is proportional to the variance $\E[y^2] = \sum_{ijk} a_{ij} v_j^1 a_{ik} v_k^1 \Cov(x_j, x_k) \gtrapprox c_0 \lambda_1^2$, where $c_0$ is the cross-correlation threshold used during the construction of functional networks. The sensing precision is thus proportional to $\lambda_1^2$, which would equal $\lambda_1 = 1 + \lra{k}$ if functional networks were random, but increases to $\lambda_1 = 1 + \lra{k} + \lra{k}/r$ because of heterogeneity. Accordingly, cell-cell communication imparts a sharp glucose-sensing acuity to \bc{-}{} collectives.

\begin{comment}
\noindent\textcolor{red}{
$x_i(t)$ is the $i$th normalized signal (i.e., zero mean and unit variance).\\
$y_i(t)=\sum_{j\in\Omega_i}{v_j^1x_j(t)}=\sum_{j}{a_{ij}v_j^1x_j(t)}$ is the signal that the $i$th node integrates from its neighbors. The summation across neighbors $j\in\Omega_i$ is replaced with the summation across the $i$th row of the adjacency matrix because $a_{ij}=1$ only if the $j$th node is the $i$th node's neighbor, otherwise $a_{ij}=0$.\\
$\mathbf{v}^1$ is the row-eigenvector corresponding to the maximum eigenvalue $\lambda_1$ such that $\sum_j{a_{ij}v_j^1=\lambda_1v_i^1}$. Also, $\mathbf{v}^1(\mathbf{v}^1)^\mathrm{T}=\sum_i{(v_i^1)^2}=1$.\\
$\E[\cdot]$ is an expectation operator that averages across time and across the whole network, i.e., it eliminates the dependence on time $t$ and index $i$.\\
$\E[y^2]=\sum_t{\sum_i{(y_i(t))^2}}$\\
$=\sum_t{\sum_i{(\sum_{j}{a_{ij}v_j^1x_j(t)})(\sum_{k}{a_{ik}v_k^1x_k(t)})}}$\\
$=\sum_i{ \sum_{j,k}{a_{ij}v_j^1a_{ik}v_k^1 \sum_t{x_j(t)x_k(t)}} }$\\
$=\sum_i{\sum_{j,k}{a_{ij}v_j^1a_{ik}v_k^1\Cov(x_j,x_k)}}$\\
$\gtrapprox{}c_0\sum_i{(\lambda_1v_i^1)(\lambda_1v_i^1)}=c_0\lambda_1^2\sum_i{(v_i^1)^2}=c_0\lambda_1^2$
}
\end{comment}

The example on glucose-sensing acuity demonstrates a physiological advantage of heterogeneous functional networks over homogeneous ones. This advantage would, however, come at a cost of vulnerability to hub-node failures~\cite{albert2000error} if heterogeneity were imprinted in the underlying physical network of \bc{ }{s}. Our results instead strongly favor an interpretation by which influential cells materialize endogenously within \bc{-}{} collectives, giving rise to influence hierarchies that are autopoietic, both in the self-producing and self-maintaining sense of this term. All that is needed for hub cells to emerge among identical peers is cell-cell communication, and should a hub fail, there is no obstacle for another one to re-emerge as long as communication remains feasible. It would thus seem that in \bc{-}{} collectives, biology has managed to create an ultra-robust architecture that is physiologically advantageous as well. This is an impressive feat that could perhaps inspire thinking about the design and engineering of next-generation critical infrastructure.

\begin{center}
--\,--\,--\,--\,--
\end{center}
\vspace{1mm}
\textbf{Acknowledgements.} DK, AS, and MSR received financial support from the Slovenian Research Agency (research core funding program no.\ P3--0396 and projects no.\ N3-0048, no.\ N3-0133 and no.\ J3-9289). MSR was further supported by the Austrian Science Fund / Fonds zur F{\"o}rderung der Wissenschaftlichen Forschung (bilateral grants I3562--B27 and I4319--B30). MJ received financial supported from the Japan Society for the Promotion of Science (JSPS) KAKENHI (grant no.\ JP 20H04288) as a co-investigator. BP received financial support from the Slovenian Research Agency (project no.\ J5-8236), the University of Zagreb's project Advanced methods and technologies in Data Science and Cooperative Systems (DATACROSS; ref. KK.01.1.1.01.009), and the University of Rijeka. PH received financial support from JSPS KAKENHI (grant no.\ JP 18H01655).

\noindent\textbf{Author contributions.} All authors contributed substantially to all aspects of the study.

\noindent\textbf{Conflict of interest.} The authors declare no conflict of interest, financial or otherwise.

\bibliography{refs}{}
\bibliographystyle{apsrev4-2}

\end{document}